\documentclass[aps,prl,twocolumn,groupedaddress,showpacs]{revtex4-1}
\usepackage{graphicx}
\newcommand{\be}{\begin{equation}}
\newcommand{\ee}{\end{equation}}
\newcommand{\bea}{\begin{eqnarray}}
\newcommand{\eea}{\end{eqnarray}}

\begin{document}

\title{Anomalous Behavior of Dark States in Quantum Gases of $^6$Li}

\author{Mariusz Semczuk, Will Gunton, William Bowden, and Kirk W.~Madison}

\affiliation{
Department of Physics and Astronomy, University of British Columbia, 6224 Agricultural Road, Vancouver, British Columbia, V6T 1Z1, Canada
}

\date{\today}
\begin{abstract}

We create atom-molecule dark states in a degenerate Fermi gas of $^6$Li in both weakly and strongly interacting regimes using two-photon Raman scattering to couple fermion pairs to bound molecular states in the ground singlet and triplet potential. Near the unitarity point in the BEC-BCS crossover regime, the atom number revival height associated with the dark state abruptly and unexpectedly decreases and remains low for magnetic fields below the Feshbach resonance center at 832.2~G.  With a weakly interacting Fermi gas at 0~G we perform precision dark-state spectroscopy of the least-bound vibrational levels of the lowest singlet and triplet potentials. From these spectra, we obtain binding energies of the $v''=9$, $N''=0$ level of the $a(1^3\Sigma_{u}^+)$ potential and the $v''=38$, $N''=0$ level of the $X(1^1\Sigma_{g}^+)$ potential with absolute uncertainty as low as $20$ kHz. For the triplet potential we resolve the molecular hyperfine structure.
\end{abstract}


\maketitle

Coherent dark states~\cite{Arimondo_76} in ultracold atomic gases lie at the heart of phenomena such as electromagnetically induced transparency~\cite{RevModPhys.77.633}, slow light~\cite{Hau_Nature_1999} and coherent population transfer~\cite{RevModPhys.70.1003}; they are useful for precision spectroscopy of molecular levels~\cite{PhysRevLett.96.023203}, and in recent years such superposition states have been demonstrated in a Bose-Einstein condensate (BEC) of atoms coupled to deeply bound molecules~\cite{PhysRevLett.95.063202}. The realization of dark states (DSs) is a crucial step towards the use of the stimulated Raman adiabatic passage to coherently transfer Feshbach molecules (atom pairs in molecular state existing very near the dissociation threshold and responsible for a Feshbach resonance) to a more deeply bound molecular level~\cite{PhysRevLett.101.133005}. Thus far the experimental demonstrations of DSs have been with bosonic species, a notable exception being that of a Bose-Fermi mixture~\cite{Ni10102008}. 

Dark states can also be created in Fermi gases~\cite{PhysRevLett.99.250404,Shao-Ying_Chin_Phys_B_2012}, and they have great potential as a probe of many-body physics~\cite{PhysRevLett.99.250404,PhysRevA.80.033606,PhysRevA.73.041607,PhysRevA.83.063605} avoiding the final state effects that complicate the interpretation of rf spectra of many-body pairs~\cite{Chin20082004,PhysRevA.84.011608}. As proposed in the context of $^6$Li, molecular dark states can be used for the optical control of magnetic Feshbach resonances~(FR)~\cite{PhysRevLett.108.010401,PhysRevA.86.063625}. This method not only suppresses spontaneous scattering but also provides larger tuning of the interactions than a single frequency approach~\cite{Bauer_Nat_Phys_2009,PhysRevA.88.041601} while enabling independent control of the effective range. Creation of a molecule-molecule DS in a molecular BEC (mBEC) may open a new path towards a BEC of ground-state molecules.

In this Letter we report (a) binding energy measurements of the least-bound vibrational states $v''= 9$ and $v''=38$ of $^6$Li$_2$ in the $a(1^3\Sigma_{u}^+)$ and $X(1^1\Sigma_{g}^+)$ potentials, respectively, and (b) creation of exotic dark states in the BEC-BCS crossover regime. 

Using the dark-state spectroscopy method~\cite{PhysRevLett.96.023203}, we achieve up to a 500-fold improvement in the measurement accuracy of the binding energies of the $v''=9$ level of the $a(1^3\Sigma_{u}^+)$ potential (in comparison to Refs.~\cite{PhysRevA.55.R3299,PhysRevA.68.051403}), resolve the molecular hyperfine structure of this level and measure the binding energy of the $v''=38$ level of the $X(1^1\Sigma_{g}^+)$ potential. The reported values are field-free extrapolations with overall accuracy as low as $20$~kHz and can be used to further refine the molecular potentials of $^{6}$Li.

In the BEC-BCS crossover regime, we verify the persistence of many-body pairing in the presence of dark-state dressing and we observe an unexpected change in the dark-state character in the vicinity of the broad Feshbach resonance at 832.2~G~\cite{PhysRevLett.110.135301}. This unanticipated behavior of the dark state may indicate new physics not previously considered and may impact the feasibility of proposals for optical FR control and dark-state probes of the system. Furthermore, we demonstrate that dark-state spectroscopy enables a general method for determining the closed-channel contribution $Z$ to the wave function of dressed molecules/pairs without relying on the details of the molecular levels involved and present measurements of $Z$ in the BEC-BCS crossover. 

We start with $2 \times 10^5$ $^6$Li atoms in an incoherent mixture of the two lowest hyperfine states (typically denoted $|1\rangle$ and $|2\rangle$) confined in a crossed optical dipole trap (CODT) at $T = 25$~$\mu$K (for details, see Ref.~\cite{PhysRevA.88.023624}). The light used to create dark states is derived from two Ti:sapphire lasers phase locked to a fiber-based frequency comb~\cite{Mills:09} resulting in a relative two-photon linewidth determined by an optical heterodyne measurement of $\Delta\nu<160$~kHz. The average frequency difference of the two laser fields is fixed and determined with an uncertainty below 1 kHz by measuring the repetition rate and offset frequency of the comb to which they are locked~\footnote{The frequency difference was independently verified by an optical heterodyne measurement up to 15~GHz}. Both beams are overlapped, have the same polarization and spatial mode, and propagate collinearly with the CODT. After the final evaporation the atoms are exposed to the photoassociation light. If a molecule in the excited state is created it either decays into a ground-state molecule or into a pair of free atoms. Both processes lead to the observed loss of atoms from the dipole trap.

\emph{Weakly interacting Fermi gas}.-- After atoms are transferred into the CODT the sample is evaporatively cooled at 300~G to $T=800$~nK~$=0.6T_F$, where $T_F$ is the Fermi temperature of a one-component Fermi gas~\footnote{For lower CODT powers we observe atom spilling due to gravitational sag in the trapping potential, an issue that cannot be resolved by magnetic levitation because at low magnetic fields the trapped states have opposite magnetic moments.}. The magnetic field is then turned off and the background magnetic field is compensated to below 20~mG. Typically, $50\times10^3$ atoms remain after the 1~s hold time.

\begin{figure}[b]
    \includegraphics[width=0.43\textwidth,angle=0]{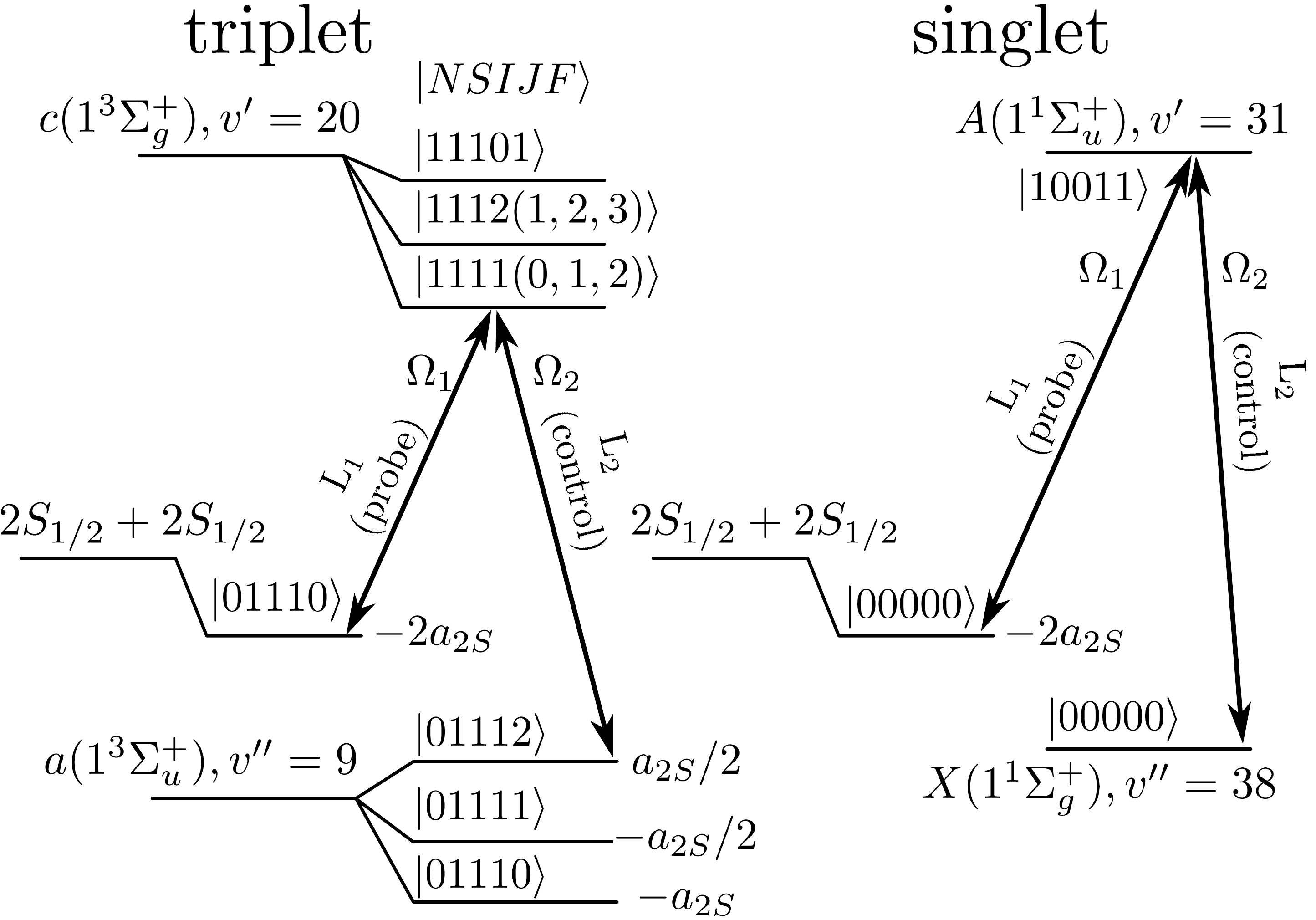}
     \caption{\label{fig:LiStateSchematic} Energy levels relevant to the experiments presented here.  At $B=0$, the initial unbound two-atom state is $|i_0\rangle = \sqrt{1/3} |00000 \rangle + \sqrt{2/3} |01110\rangle$, a linear combination of the singlet and triplet states shown lying $2a_{2S}$ below the $2S_{1/2} + 2S_{1/2}$ asymptote.  The levels are labeled with the quantum numbers $|NSIJF\rangle$, where $\vec{N}$ is the molecular rotational angular momentum, $\vec{S}$($\vec{I}$) is total electronic (nuclear) spin, $\vec{J}$ is total angular momentum apart from nuclear spin and $\vec{F}=\vec{J}+\vec{I}$. Here, $\Omega_1$ and $\Omega_2$ are Rabi frequencies of transitions driven by the lasers $L_1$ and $L_2$, respectively, and $a_{2S}=152.137$~MHz is the atomic magnetic dipole hyperfine constant of $2^2S_{1/2}$.}
\end{figure}

The initial unbound two-atom state has a spatially symmetric $s$-wave scattering wave function ($N=0$) and an antisymmetric spin state with $f_1 = f_2 = 1/2$ and total spin of $F = 0$. At $B=0$, $F$ and $m_F$ are good quantum numbers, but to explicitly show the singlet and triplet character of the initial state $|i_0\rangle$, we follow Ref.~\cite{PhysRevA.75.023612} and represent it in the molecular basis ($|NSIJF\rangle$) shown in Fig.~\ref{fig:LiStateSchematic}, such that $|i_0\rangle = \sqrt{1/3} |00000 \rangle + \sqrt{2/3} |01110\rangle$.

\begin{figure}[b]
\includegraphics[width=0.4\textwidth,angle=0]{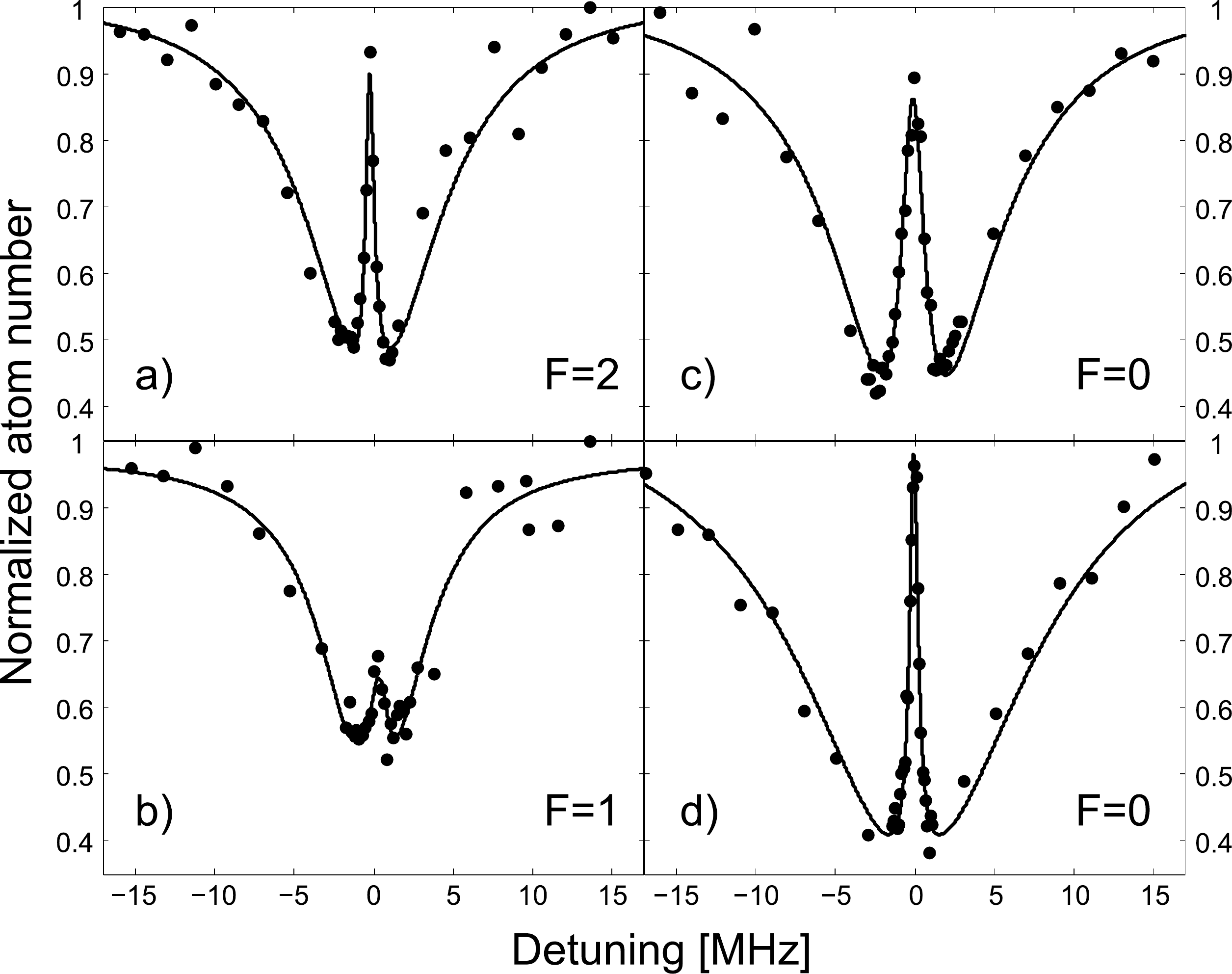}
\caption{\label{fig:ds_zero_G} Dark states in a Fermi gas at 0~G. The atom number is plotted versus the detuning of the probe frequency from the two-photon resonance, where the photon energy difference $h(\nu_2-\nu_1)$ equals the difference between the energy of the initial free atom state and a bound molecular level and the atom number exhibits a maximum revival. (a) - (c) Corresponding different molecular hyperfine levels ($F''=2,1,0$) of the $v''=9$, $N''=0$ level in the $a(1^3\Sigma_{u}^+)$ potential. We were not able to find parameters that would improve the revival of $F''=1$ to above 50\%. Spectrum (d) corresponds to the $v''=38$, $N''=I''=F''=0$ level of the $X(1^1\Sigma_{g}^+)$ potential.}
\end{figure}

The illumination time and light intensity $I_1$ of laser $L_1$ with frequency $\nu_1$ is chosen such that after 1~s $L_1$ induces a loss of 40$\%$-80$\%$ for intensities 100-500~W/cm$^2$ corresponding to Rabi frequencies $\Omega_1\ll 1$~kHz. The intensity of $L_2$ (20-200~W/cm$^2)$ is chosen so that the corresponding Rabi frequency $\Omega_2\gg \Omega_1$ but also low enough ($\Omega_2<1$~MHz) to avoid inducing large Autler-Townes splitting of the excited state. The frequency of $L_2$, $\nu_2$ is set to match the $v''=9\leftrightarrow v'=20$ ($v''=38\leftrightarrow v'=31$) transition between the triplet(singlet) levels using the frequencies initially determined by two-color photoassociation spectroscopy, as in Ref.~\cite{PhysRevLett.74.1315}. The choice of the excited levels is based on empirical evidence that their ac Stark shift due to the field from laser $L_1$ is small in comparison with other levels measured in Refs.\cite{PhysRevA.87.052505,PhysRevA.88.062510}. The frequency $\nu_1$ is scanned over a range that induces atom loss due to the single color photoassociation of free atoms to $v'=20$ or $31$ vibrational levels. With $L_2$ on, dark-state spectra are observed as shown in Fig.~\ref{fig:ds_zero_G}. When the two-photon resonance condition is fulfilled an atom-molecule dark state is created and the loss induced by $L_1$ is almost completely suppressed for all levels but $|g_1\rangle=|v''=9,N''=0,F''=1\rangle$ where the suppression is only partial [Fig.~\ref{fig:ds_zero_G}(b)]. The poor revival is unlikely to result from the finite collisional lifetime of $|g_1\rangle$ as it would have to be unreasonably short, $\sim$1~$\mu$s.

To determine the ac Stark shifts induced by the lasers $L_1$, $L_2$ and the dipole trap, the DS spectra are measured for several dipole trap depths with the control(probe) beam intensity 50(350)~W/cm$^2$. At the lowest trap depth additional spectra are measured when both the control and the probe beam intensities are varied. A three-level model (as in e.g., Ref.~\cite{PhysRevLett.95.063202}) is used to fit each data set to extract the two-photon resonance position. The resulting values for each set of beam intensities are used to extract the field-free binding energies. 

\begin{table}
\caption{\label{tab:binding_energies}Experimentally measured binding energies of the least-bound vibrational levels of the $a(1^3\Sigma_{u}^+)$ and $X(1^1\Sigma_{g}^+)$ potentials of $^{6}$Li$_{2}$. The frequency difference $\nu_2-\nu_1$ is extracted from the dark-state spectra and corresponds to the energy difference between the initial and the final state. The quoted uncertainties represent the statistical uncertainties on the fits. The systematic uncertainty is below 1~kHz.}
\begin{ruledtabular}
\begin{tabular}{cccc}
$v''$&$F''$&\multicolumn{1}{c}{$\nu_2-\nu_1$ [GHz]}&\multicolumn{1}{c}{Binding energy\footnote{The initial free atomic state is $2a_{2S}$ below the hyperfine center of gravity of the $2S_{1/2}+2S_{1/2}$ asymptote therefore the binding energy is computed by adding $2a_{2S}=304.274$~MHz to the measured frequency difference.} [GHz]}\\
\hline 
\noalign{\smallskip}
\multicolumn{4}{c}{$X(1^1\Sigma_{g}^+)$}\\\noalign{\smallskip}
\hline
\hline\noalign{\smallskip}
$38$ & 0 & 1.321671(21) & 1.625945(21)\\\hline\noalign{\smallskip}
\multicolumn{4}{c}{$a(1^3\Sigma_{u}^+)$}\\\noalign{\smallskip}
\hline
\hline\noalign{\smallskip}
 & 2 & 24.010649(46) & 24.314923(46)\\
 9 & 1 & 24.163035(105)& 24.467309(105)\\
 & 0 & 24.238372(54) & 24.542646(54)\\
\end{tabular}
\end{ruledtabular}
\end{table}
The dark-state spectroscopy is a differential measurement; therefore only the frequency difference of the lasers $L_1$ and $L_2$ needs to be known precisely. The results are summarized in Table~\ref{tab:binding_energies}. We note that the molecular hyperfine splittings inferred from Table~\ref{tab:binding_energies} differ from theory~\cite{PhysRevA.51.R4333,PhysRevA.53.3092} by more than can be explained by the uncertainties of our measurements.

We observe a complete suppression of loss on two-photon resonance even for illumination times close to 2~s. To the best of our knowledge, the longest times reported in experiments with other species rarely exceed a few tens of ms. To create dark states we turn the laser fields on and off in the following order: $L_2$~on, $L_1$~on, $L_1$~off, $L_2$~off, where the turn on or off times are $\ll 1$~$\mu$s. The turn on or off of $L_1$ is adiabatic for most of the cases; however,when the turn on or off of $L_1$ is nonadiabatic, we expect a loss of atoms due to the projection of the initial state onto the bright state. When $\Omega_2 \gg \Omega_1$, this projection loss is $2(\Omega_1 / \Omega_2)^2$. The second loss source is due to the projection of the atom-molecule state $|\mathrm{AM}\rangle$ onto the new bright state $|\mathrm{BS}\rangle$

\begin{equation}\label{eq:ds_loss_lasers}
|\langle \mathrm{AM}|\mathrm{BS}\rangle|^2 = \frac{2\Omega^2_1\Omega^2_2(1-\cos(\phi))}{(\Omega^2_1+\Omega^2_2)^2}
\end{equation}
caused by a nonadiabatic phase jumps of the lasers. Here, $\phi$ is the new relative phase between two lasers after the jump. If we model the phase jitter of the lasers as producing a nonadiabatic jump of the relative phase by an average $\phi=\pi/2$ every dephasing time $\tau \sim 1/\Delta\nu = 6.2$~$\mu$s, after 350~ms this projection, for $\Omega_2/\Omega_1 > 1000$, will lead to an atom loss of less than our signal-to-noise value. This simple model significantly underestimates the observed lifetime.

\emph{BEC-BCS crossover regime}.-- When the final evaporation is done at magnetic fields close to the broad FR at 832.2~G, pairs form, here referred to as Feshbach molecules below (where a two-body bound state exists) and BCS-like pairs above resonance (where pairing is a many-body phenomenon). In this case, the initial Feshbach-dressed molecule or pair state is $|i_{\mathrm{c}}\rangle = \sqrt{Z} |g_{\mathrm{closed}}\rangle +
\sqrt{1-Z}|g_{\mathrm{open}}\rangle$, where the open channel has almost a pure triplet character and is strongly coupled to the closed channel molecular state responsible for the wide FR, $|g_{\mathrm{closed}}\rangle \equiv (2 \sqrt{2} |00000\rangle - |00200\rangle)/3$, a linear combination of the $I=0\;\mathrm{and}\;2$ ($v''=38$) singlet states \cite{PhysRevA.75.023612, PhysRevA.86.063625}. To form dark states, we use the singlet levels $|e\rangle$ ($v'=31$,  $|00011\rangle$) and $|g\rangle$ ($v''=37$, $|00011\rangle$). In the crossover regime the $v''=38$ singlet level is nearly degenerate with the entrance (open) channel; therefore, another level $|g\rangle$ in the $X(1^1\Sigma_{g}^+)$ potential is required, here chosen to be $v''=37$. The apparent binding energy change of $v''=37$ at nonzero magnetic fields is caused by the shift of the initial free atom state $|i_{\mathrm{c}}\rangle$. The observed energy differences between $|i_{\mathrm{c}}\rangle$ and $|g\rangle$ are summarized in Table~\ref{tab:binding_energies_Bfield}.

\begin{figure}
\includegraphics[width=0.48\textwidth,angle=0]{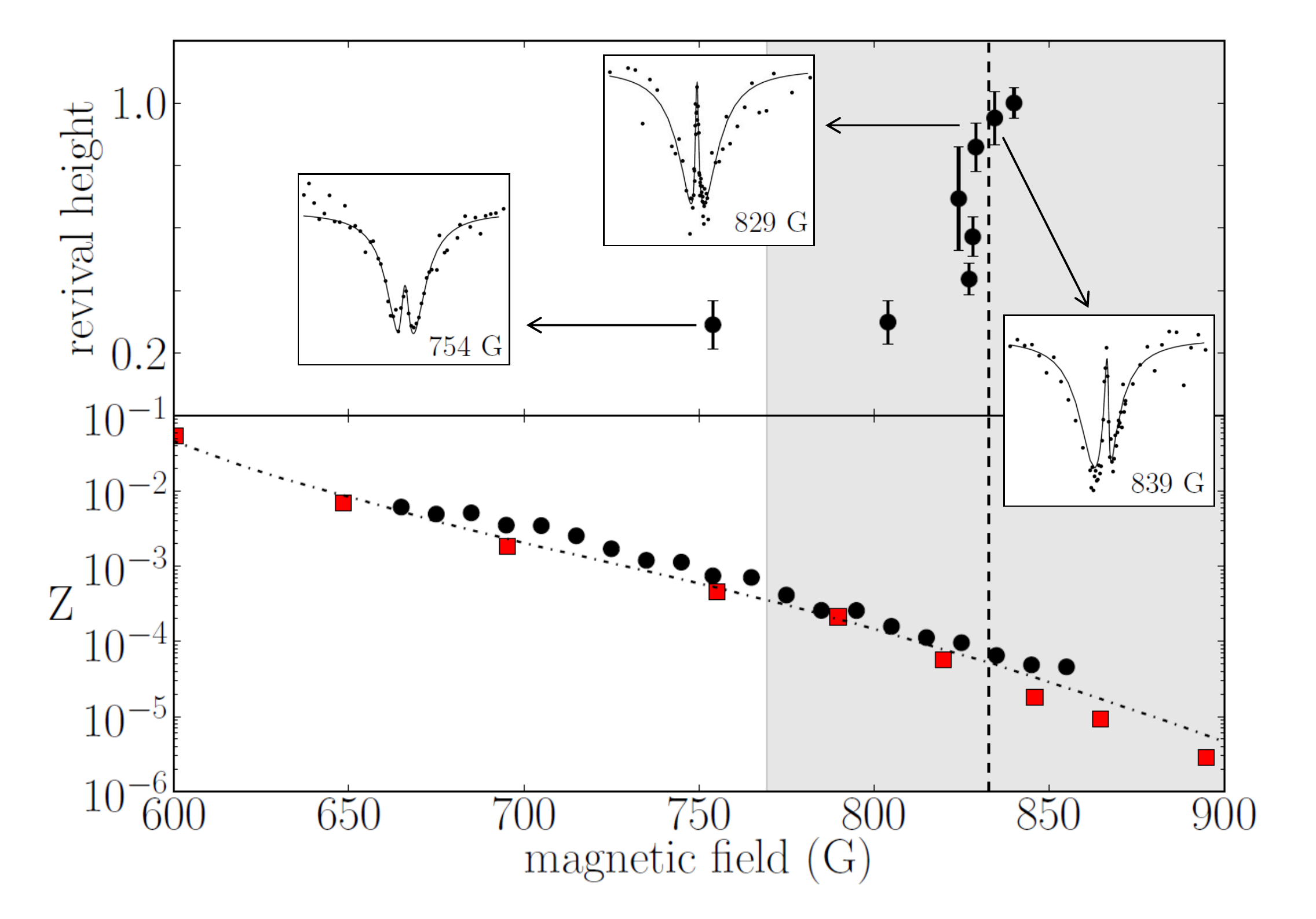}
\caption{\label{fig:revival_height}(color online.) (top) The revival height of the dark state defined as a ratio of the suppression amplitude to the loss amplitude, measured on both sides of the Feshbach resonance (dashed vertical line). The states involved are: Feshbach molecules or BCS-like pairs ($|i_{\mathrm{c}}\rangle$),  $v'=31$ in the $A(1^1\Sigma_{u}^+)$ excited molecular potential ($|e\rangle$) and $v''=37$ in the $X(1^1\Sigma_{g}^+)$ ground-state potential ($|g\rangle$). For magnetic fields above 829~G we observe near complete revival on the two-photon resonance. Insets: spectra at selected magnetic fields. (bottom) Full circles represent the probability $Z$ of the dressed molecule to be in the closed channel. Squares (red) show $Z$ measured in Ref.~\cite{PhysRevLett.95.020404} where the excited state $|e\rangle$ is $v'=68$ [$A(1^1\Sigma_{u}^+)$ potential]. Dash-dotted line shows theoretical prediction for $Z$ taken from Ref.~\cite{PhysRevA.74.053618}. The gray region corresponds to $k_F|a|>1$. For these data, $T/T_F = 0.4 \pm 0.15$, $E_F/ h = 11$~kHz, $\Omega_1 = 200$~kHz, and $\Omega_2 \simeq 2$~MHz.}
\end{figure}

Evaporation below 832.2~G results in a mBEC. For intensities $I_1=0.045$-$10$~kW/cm$^2$ and $I_2=0.040$-$0.3$~kW/cm$^2$, and illumination times 5~$\mu$s to 100~ms (short times correspond to large $I_1$) we observe dark-state features that show only partial loss suppression on the two-photon resonance (Fig.~\ref{fig:revival_height}(top)). Both $\Omega_1$ and $\Omega_2$ as well as the exposure time have been varied over a range where a full revival is expected. Nevertheless, we always observe revival heights below 50\%, even for temperatures above the mBEC critical temperature.

When a degenerate Fermi gas is prepared above the resonance such that there are BCS-like pairs present, we observe DS features corresponding to a coherent superposition of these pairs and molecules in the $|g\rangle$ level of the singlet potential [Fig.~\ref{fig:revival_height}(top)]. 
In order to confirm that these dark states can be created without disrupting the many-body pairing physics, the frequencies $\nu_1$ and $\nu_2$ are set to match the two-photon resonance condition where negligible dark-state tuning of the scattering length is expected and rf spectroscopy with 200 ms long rf pulses is performed, revealing a spectrum consistent with the presence of pairing~\cite{Chin20082004,PhysRevA.88.023624}. Here, $\Omega_1/2\pi = 5$~kHz and $\Omega_2 / \Omega_1 \sim 1000$; thus, the BCS-like pairs were only weakly dressed with the $|g\rangle$ level, and the many-body interactions were, thus, negligibly perturbed.

\begin{table}[b]
\caption{\label{tab:binding_energies_Bfield}Energy difference between the initial atomic state and the final singlet state $|g\rangle$ ($v''=37$, $|00011\rangle$) measured at selected magnetic fields. The uncertainties are conservatively estimated to be 1~MHz.} 
\begin{ruledtabular}
\begin{tabular}{ccccc}
$B$ [gauss]&0&754&804&840\\\hline\noalign{\smallskip}
$\nu_2-\nu_1$ [GHz]&58.260&56.364&56.225&56.124\\
\end{tabular}
\end{ruledtabular}
\end{table}
The revival height on the two-photon resonance is studied in the crossover regime using an exposure time of 40~$\mu$s. Figure~\ref{fig:revival_height}(top) shows that the revival height changes abruptly for fields below 829~G. This change, coincidentally, occurs at magnetic fields where the two-body bound state is present. However, we observe a near full revival also below the Feshbach resonance center at 832.2~G~\cite{PhysRevLett.110.135301}.
This abrupt change in the revival height was unexpected, and we, therefore, independently checked that
$\Omega_1$ is continuous in this regime by performing single-color photoassociation of $|i_{\mathrm{c}}\rangle$ to $|e\rangle$. Only the $I=0$ part of the closed channel contributes to photoassociation to $|e\rangle$; therefore

\bea
&&\Omega_1(B) =  \langle i_{\mathrm{c}} | \vec{d}\cdot \vec{E} | e \rangle = \langle g_{\mathrm{closed}} | \vec{d}\cdot \vec{E} |10011 \rangle_{v'=31}
\\
\nonumber
& &=  \sqrt{Z} \sqrt{\frac{8}{9}} \langle 00000_{v''=38}  | \vec{d}\cdot \vec{E} |10011\rangle_{v'=31} = \sqrt{\frac{8Z}{9}} \Omega_0.
\eea
We determine $\Omega_0$ experimentally from a fit of the measured dark-state spectra at $B=0$ shown in Fig.~\ref{fig:ds_zero_G}(d). In that spectra, $\Omega_0$ plays the role of the bound-to-bound coupling $\Omega_2$. We observe that $Z \equiv \left(9/8\right)\left(\Omega_1/\Omega_0\right)^2$ [shown in Fig.~\ref{fig:revival_height}(bottom)] is continuous in the region where the dark-state revival changes abruptly. Our observations are consistent with those reported in Ref.~\cite{PhysRevLett.95.020404}; however, our determination of $Z$ does not rely on the calculation of $\Omega_0$ and, thus, is independent of a theoretical model for the molecular potentials.

In summary, we have demonstrated dark states in degenerate gases of $^6$Li and used them to measure the binding energy of the least-bound vibrational levels in the $a(1^3\Sigma_{u}^+)$ and $X(1^1\Sigma_{g}^+)$ potentials with accuracy up to 500 times better than previous measurements~\cite{PhysRevA.55.R3299,PhysRevA.68.051403}. Using $\Omega_2$ extracted from the field-free measurements of the $v''=38$ level we directly measure the closed channel contribution to the dressed molecules or pairs near the broad FR, a method that is independent of the details of the molecular potentials involved. In addition, we observe that under certain conditions the dark-state lifetime, as determined by the revival peak, can far exceed the two-photon coherence time of the lasers producing it. Finally, our measurements show an unexpected, abrupt change in the revival height of the dark-state features in the BEC-BCS crossover. 

\begin{acknowledgments}
We gratefully acknowledge M.G.~Raizen and T.~Momose for the use of the Ti:sapphire lasers.  We also thank X. Li and E. Shapiro for many helpful discussions.  The authors acknowledge financial support from the Natural Sciences and Engineering Research Council of Canada (NSERC / CRSNG) and the Canadian Foundation for Innovation (CFI). 
\end{acknowledgments}


\providecommand{\noopsort}[1]{}\providecommand{\singleletter}[1]{#1}%

\end{document}